\documentclass[%
floatfix,
aps,
jor,
amsmath,amssymb,
reprint,%
]{revtex4-1} 

\usepackage{amsmath,amssymb}
\usepackage{graphicx}
\usepackage{dcolumn}
\usepackage{bm}
\usepackage{amsmath}
\usepackage{subfig}
\usepackage{floatrow}
\usepackage{hyperref}
\usepackage[autostyle]{csquotes}
\usepackage{microtype}
\usepackage{comment}

\hypersetup{
    colorlinks=true,
    citecolor=blue,
    linkcolor=blue,
    filecolor=magenta,      
    urlcolor=cyan,
    pdftitle={Overleaf Example},
    pdfpagemode=FullScreen,
    }

\begin{document}


\title[Sample title]{\textbf{Entropic pulling and diffusion diode in an It\^o process}
}

\author{Mayank Sharma}
 \email{mayank.sharma@students.iiserpune.ac.in}
\author{A. Bhattacharyay}%
 \email{a.bhattacharyay@iiserpune.ac.in}
\affiliation{ 
Indian Institute of Science Education and Research, Pune, India
}%

\date{\today}

\begin{abstract}
 Biological environments at micrometer scales and below are often crowded, and experience incessant stochastic thermal fluctuations. The presence of membranes/pores, and multiple biological entities in a constricted space can make the damping/diffusion inhomogeneous. This effect of inhomogeneity is presented by the diffusion becoming coordinate-dependent. In this paper, we analyze the consequence of inhomogeneity-induced coordinate-dependent diffusion on Brownian systems in thermal equilibrium under the It\^o's interpretation. We argue that the presence of coordinate-dependent diffusion under It\^o's formulation gives rise to an effective diffusion potential (and, equivalently an entropy) that can have substantial contribution to system's transport. This emergent force when looked at as of entropic origin it provides a physical basis of the notion of the entropic pulling postulated in the context of working of some biological systems.
\end{abstract}

\keywords{Suggested keywords}
\maketitle





Many biological systems operate at low Reynolds numbers, where viscous forces are of paramount importance. Viscous/damping forces dominate over inertial forces and govern the mechanics of such biological systems in this regime \cite{purcell1977life, berg1977physics, nachtigall1981hydromechanics, purcell1997efficiency,cohen2010swimming}. Coordinate-dependent damping/diffusion is recently identified to have crucial role to play in the functioning of various biological systems \cite{foster2018probing, neupane2017direct, cellmer2008measuring, faucheux1994confined, lin2000direct, noguchi2016state} and a lot remains to be explored in this direction. Earlier experiments by Faucheux et al, and others have proven the existence of coordinate-dependent damping \cite{faucheux1994confined, lin2000direct, sharma2010high} of a Brownian particle near interfaces due to hydrodynamic effects. Ample theoretical/computational studies have been focusing on understanding systems with coordinate-dependent diffusion/damping as opposed to the uniform one \cite{berezhkovskii2017communication, best2010coordinate, oliveira2010coordinate, roussel2004reaction, chahine2007configuration, bhattacharyay2019equilibrium} which by now is well understood.

Translocation of proteins across cellular membranes is a central and essential process in biological systems. Molecular chaperones are macromolecules present in cells that assist these proteins to unfold, move across membranes/channels to prevent protein aggregation and help in maintaining proper functioning and health of the cell. Broadly, three mechanisms have been proposed in the literature to explain chaperone-assisted translocation of proteins: {\it power-stroke}, {\it Brownian-ratchet} and {\it entropic-pulling} \cite{goloubinoff2007mechanism, sousa2019physics, de2006hsp70, sousa2006keep, rohland2022conformational, tomkiewicz2007pushing, de2016hsp70, ganesan2019structural, yu2023structural, hwang2019structural, craig2018hsp70}. In the power-stroke mechanism, the linkage of incoming protein with the chaperone and assistance from ATP hydrolysis induces a conformation change within chaperone that drives the protein in forward direction through the channel/pore. The Brownian-ratchet mechanism is a biased diffusion model effectively based on the idea that while passing through a pore/channel, the large size of the chaperone only allows the Brownian motion of protein-bound chaperone in one direction. 

On the contrary, the entropic pulling mechanism is a thermodynamic description based on the idea that the system would try to move in a direction so as to attain a higher entropy configuration \cite{roos2014entropic, neumann1980entropic, Rukes2024.02.06.579217}. It is generally understood in existing literature that, in chaperone-assisted protein translocation, the tethering of protein to chaperone in the vicinity of the pore/membrane corresponds to a low entropy state. The chaperone-linked protein system via collisions with the membrane/pore generates a net effective pulling force that takes it away from pore/membrane to a region of more freedom in conformations. This constitutes an effective force transitioning to a higher entropy state.

Entropic-pulling of a meso-scale object from open space to a narrower one is beyond any scope under homogeneous diffusion and that makes such a phenomenon counter-intuitive. In the present paper, we are going to illustrate effects of a new source of entropy arising out of coordinate-dependent diffusion \cite{bhattacharyay2023equilibrium} under It\^o's convention. The presence of this entropy eventually could give rise to an entropic-force which would drag a Brownian particle or a mesoscopic object towards a region of higher damping despite the region of higher damping being a relatively constricted space. In general, coordinate-dependent diffusion is a hydrodynamic effect, where a Brownian particle experiences enhanced damping near an interface or wall than that it undergoes in the bulk. This makes the particle spend more time near a wall compared to that in the bulk (in an It\^o process) and this phenomenon can be understood in terms of an inhomogeneity-entropy arising out of coordinate dependence of diffusion. We demonstrate working of this entropic-force, in the present paper, in two classes of phenomenon. These two phenomena are (1) {\it diffusion-diode} and (2) {\it chain-pulling} through a pore. 

In the absence of coordinate-dependent diffusion, diffusive transport is always driven by concentration gradients. However, in the presence of diffusivity gradient, {\it in an It\^o process}, the gradient can force a diffusive transport even against concentration gradient. If one imagines in such a process that the concentration gradient is representing a driving field for current then, until this driving force overcomes the barrier of the diffusion gradient it will not be able to cause substantial transport in its direction. This, however, is not the case when the concentration gradient and the diffusion gradient are in the same direction. The operating principle of a diffusion-diode would be based on cooperation/opposition of these two gradients and can be realized in the presence of coordinate dependence of diffusion as an It\^o process.

In equilibrium, diffusivity and damping are inversely related by fluctuation-dissipation relation locally in space which becomes a global relation for uniform diffusion and damping over space. Considering, therefore, local equilibrium scenario, diffusivity decreases near a wall due to increase in damping due to hydrodynamic effects. This decrease in diffusivity creates a higher entropy region near a wall compared to that in the bulk and that picture is the essential ingredient of an It\^o process. Therefore, there always be a tendency in the diffusive transport of macro-molecules to make them get attracted towards a wall/pore near which the diffusivity is smaller in magnitude. 
\par{}
Due to this same effect, a chain-like macro-molecule left to meander under equilibrium thermal fluctuations would eventually move from a wider region of space to a nearby narrower one under the action of coordinate-dependent diffusion. To our knowledge, possibility of occurrence of this phenomenon has not been paid attention to due mostly to not having considered the It\^o-process as an equilibrium scenario for coordinate-dependent diffusion. In this paper, we are not going into a discussion on the so-called It\^o-vs-Stratonovich controversy, but, are solely interested in looking at coordinate dependent diffusion as an It\^o process and present some striking consequences of this stochastic process on transport. These are novel phenomena possible under coordinate-dependent diffusion which might hold the key to understanding complex biological systems in a substantially new way.

The paper is organized in the following manner: the first section revisits basics of an It\^o-process using Fokker-Planck equation and demonstrates how a force of entropic origin emerges therein. Subsequently, we introduce a simple model to demonstrate the working of a {\it diffusion-diode}. This is followed by a section on {\it entropic-pulling} based translocation of a chain-like molecule through a pore. The paper is concluded by discussing implications of inhomogeneity-entropy arising due to coordinate-dependent damping/diffusion in the context of biological systems.

\section{Coordinate-dependent diffusion and It\^o process}
Stochastic modeling has played an eminent role in helping biologists and physicists understand various natural phenomena \cite{bressloff2014stochastic, allen2010introduction, phillips2012physical, alon2019introduction} like the directed motion of molecular motors, the spread of epidemics, gene regulation etc. The evolution of such stochastic systems are often understood by expressing the dynamics as a partial differential equation in the presence of a thermal noise. In one spatial dimension $x$ when diffusivity is coordinate dependent $D(x)$, an It\^o-process is governed by the following Fokker-Planck equation \cite{peliti2021stochastic,bhattacharyay2019equilibrium}:

\begin{eqnarray*}
    \frac{  \partial }{\partial t}p(x,t) &=& -\frac{ \partial}{\partial x}[\nu(x)p(x,t)] + \frac{\partial^{2}}{\partial x^{2}}[D(x)p(x,t)].
\end{eqnarray*}

where $p(x,t)$ is the probability density of the particle at position $x$ and time $t$, $\nu(x)=\frac{F(x)}{\Gamma(x)}$ is the drift velocity (coefficient) and $D(x)$ is the diffusion coefficient. The deterministic (conservative) force on the particle being $F(x)$, the damping coefficient is denoted by $\Gamma(x)$ which, in general, is related to the diffusivity as $\Gamma(x)=\frac{k_BT}{D(x)}$ by the fluctuation-dissipation (Stokes-Einstein) relation under local equilibrium conditions. The equilibrium is maintained at temperature $T$ where $k_B$ is the Boltzmann constant. 

The Fokker-Planck equation is a continuity equation for conserved probability:
\begin{eqnarray*}
    \frac{  \partial }{\partial t}p(x,t) &=& -\frac{ \partial}{\partial x}J(x,t).
\end{eqnarray*}
where $J(x,t)$ is the probability current density, with $J(x,t)=\nu(x)p(x,t) - \frac{\partial }{\partial x}D(x)p(x,t)$. The first term here corresponds to drift current density and second term to diffusion current density{ \cite{ sattin2008fick, bengfort2016fokker,vazquez2010non, van2005applicability,peliti2021stochastic, bhattacharyay2019equilibrium,maniar2021random,sharma2020conversion,bhattacharyay2023equilibrium}. Now, for a stationary distribution in equilibrium $J^{eq}(x) =0$ (condition for detailed balance), the following equilibrium distribution $p^{eq}(x)$ results:
\begin{eqnarray*}
     p^{eq}(x) = C\frac{D_{0}}{D(x)} \exp\left(\int dx \frac{\nu(x)}{D(x)}\right).
\end{eqnarray*}
Here $C$ is a constant with dimensions of inverse length, and $D_{0}$ is the bulk diffusivity away from any interface to which $D(x)$ converge as the Brownian particle makes its excursion away from an interface.

This distribution is referred to as generalized/modified Boltzmann distribution or It\^o-distribution in \cite{bhattacharyay2019equilibrium, dhawan2024ito, bhattacharyay2020generalization, maniar2021random}. Now, the drift velocity $\nu(x)$ can be expressed in terms of potential $U(x)$ corresponding to a conservative force $F(x)$ and coordinate dependent damping $\Gamma(x)$ as : $\nu(x)=-\frac{1}{\Gamma(x)}\frac{\partial}{\partial x}U(x)$. The equilibrium distribution $p^{eq}(x)$ thus becomes \cite{sharma2020conversion, sharma2023spontaneous, bhattacharyay2019equilibrium, maniar2021random}
\begin{eqnarray}
     p^{eq}(x) = C\frac{D_{0}}{D(x)} \exp\left(-\frac{U(x)}{k_{B}T}\right).
\end{eqnarray}

 The presence of the coordinate-dependent diffusivity-dependent amplitude $D_0/D(x)$, makes all the difference and is responsible for the possibility of various interesting processes in thermal equilibrium like bath-fluctuation-driven spontaneous collective transport \cite{sharma2023spontaneous}, rectified transport of symmetry broken hetero-dimer \cite{bhattacharyay2012directed, sharma2020conversion} etc. The legitimacy of the distribution (1) as an equilibrium distribution could be illustrated by rewriting $p^{eq}(x)$ as 
\begin{equation}
p^{eq}(x)= C \exp\left(-\frac{1}{k_{B}T}\left[ U(x)-k_{B}T\log\frac{D_{0}}{D(x)}\right]\right),
\end{equation}
where interpretation of the above expression could be that, the presence of coordinate dependence of diffusion generates a diffusion potential $U_{D(x)}=-k_{B}T \log\frac{D_{0}}{D(x)}$, which corresponds to a force $f_{D(x)}=-\frac{\partial}{\partial x}U_{D(x)} =-k_{B}T \frac{d}{dx}\log D(x)$.

This force supports the motion in the direction of decreasing diffusion. This is the Molecular-kinetic/tree description in the context of the paper by Sousa and Lafer \cite{sousa2019physics}. Another way to interpret this is the thermodynamic description \cite{sousa2019physics,bhattacharyay2023equilibrium}, where one interprets the presence of position-dependent diffusion responsible for the emergence of additional entropy. The term $\frac{D_{0}}{D(x)}$ can be interpreted as the dimensionless density of states arising in the presence of coordinate-dependent diffusion \cite{bhattacharyay2023equilibrium}. The corresponding force $f_{E}(x) = -\frac{d\mathcal{F}}{dx} = T \frac{dS}{dx}$. Here $\mathcal{F}$ corresponds to free energy. This in our case means $f_{E}(x)=- k_{B}T \frac{d}{dx} \log D(x)$. This force could be imagined to result in the entropic-pulling mechanism. In the following sections, we are going to demonstrate the effects of It\^o's distribution by demonstrating the possibility of a diffusion diode and thermal fluctuations induced stochastic pulling of a chain-like molecule from a wider region to a narrower region. In both of these applications shown, the basic physics is dictated by the It\^o's distribution.

\section{DIFFUSION DIODE}
Consider a model consisting of $N$ Brownian particles confined inside a three-dimensional box in the presence of a heat bath in equilibrium at temperature $T$. Brownian particles interact with each other through excluded volume interaction in the vicinity of one another. Overdamped stochastic differential equations 
\begin{eqnarray}
    \frac{  d{ \mathbf { r}_{i}  } }{dt} &=& \frac{ \mathbf{F}_{i}}{\Gamma_{i}} +\sqrt\frac{2k_{B}T}{ \Gamma_{i}} {\boldsymbol{\eta}}_{i}(t),
\end{eqnarray}
 govern such a system where $\boldsymbol{r}_{i}$ is the position coordinate of $i^{th}$ particle. $\boldsymbol{F}_{i}$ is the instantaneous force amounting to the total contribution of excluded volume interaction and the confinement felt by the $i^{th}$ particle, in case it tries to cross the boundary. $k_{B}T$ represents the thermal energy scale of the heat bath at temperature $T$. $\Gamma_{i}$ is coordinate-dependent damping coefficient assigned to the $i^{th}$ particle. Stochastic noise $\boldsymbol{\eta}_{i}(t)$ felt by $i^{th}$ particle is a three-component Gaussian white noise vector, which in Cartesian coordinate representation is $\boldsymbol{\eta}_{i}(t) = {\eta}^{x}_{i}(t) \hat{{x}} +{\eta}^{y}_{i}(t) \hat{{y}} + {\eta}^{z}_{i}(t) \hat{{z}}$. Each component represents Gaussian white noise of zero mean and a unit strength. None of these components are cross-correlated i.e. $\langle \eta^s_{i}(t) \rangle$ = 0 and $\langle \eta^s_{i}(t_{1})\eta^{s^\prime}_{j}{(t_{2})} \rangle = \delta_{ij}\delta_{ss^\prime}\delta(t_1-t_2)$, with $i$, $j$ $\in$ $\{1,2,\ldots\,N\}$   and $s,s^\prime \in \{x,y,z\}$.\\

 We set up a box in the first octant, with the origin (0, 0, 0) coincident on one of the corners. All distances are measured from the origin. Length of the box in $y$ and $z$ directions are equal by construction. We introduce a global coordinate-dependence of damping in the box in this model. This is achieved by dividing the box into three cuboidal regions: $R_{1}$, $R_{2}$ and $R_{3}$ along $x$-axis, where regions $R_{1}$ and $R_{3}$ are geometrically identical. The damping profile is such that the damping strength is almost constant in $R_{1}$, increasing in region $R_{2}$ till it reaches region $R_{3}$, where it again saturates to another almost constant value. The damping profile chosen is a monotonically increasing, continuous, and differentiable function. \\
 \par{}
 We study two cases under such a scenario:- Case I: $N$ Brownian particles are randomly placed in $R_{1}$ at the start of the simulation, and Case II: $N$ Brownian particles are randomly placed in $R_{3}$ at the start of the simulation. We allow the system of particles to evolve numerically as an It\^o-process and compare the dynamics in both cases. \\  

\begin{figure}
 \includegraphics[scale=0.35]{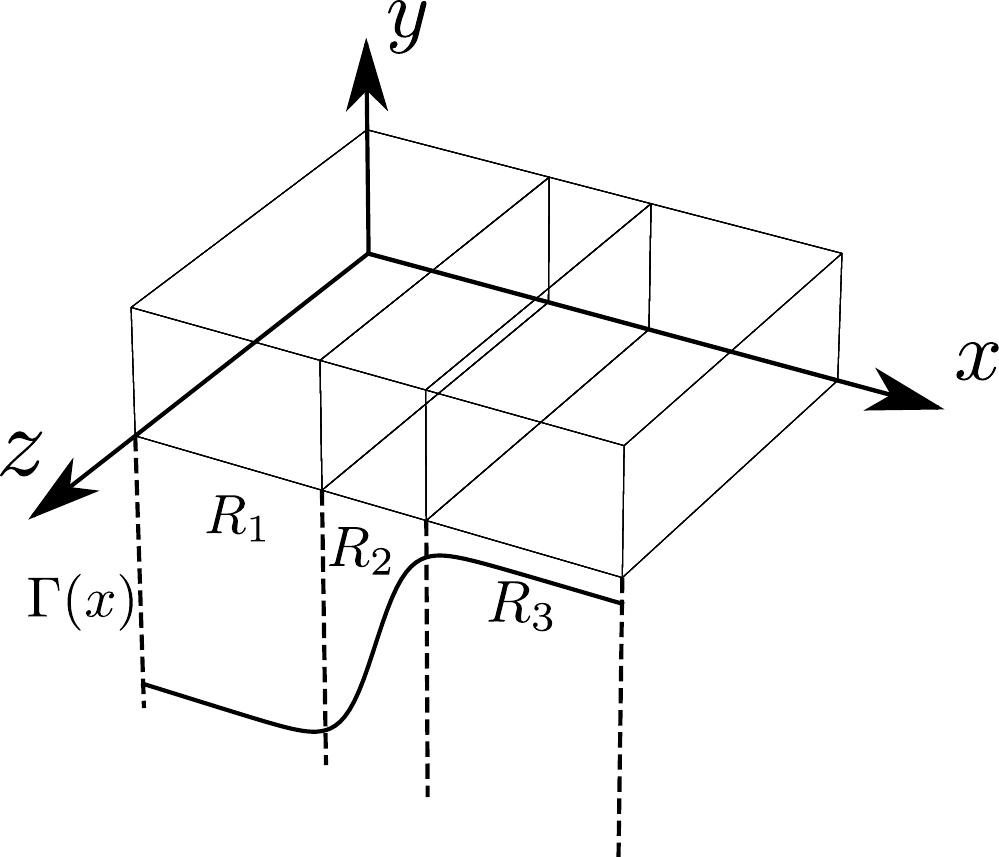}
 \caption{ \footnotesize   Set up of the model (not drawn to scale)}
 \label{fig_1}
\end{figure}

  The excluded volume interaction between different particles $i$ and $j$ is modeled by repulsive harmonic potential $V_{ij}$.
\begin{eqnarray}\nonumber
    V_{ij} =
  \begin{cases}
                 \frac{1}{2}\kappa (|\mathbf { r}_{i}-\mathbf { r}_{j}|-r_{l})^{2}& \text{if $|\mathbf { r}_{i}-\mathbf { r}_{j}|<r_{l},$} \\\\
                0 & \text{if $|\mathbf { r}_{i}-\mathbf { r}_{j}|\geq r_{l},$}
  \end{cases}
\end{eqnarray}\\
\par
where, $\kappa$ measures the strength of repulsion, $r_{l}$ is the proximity scale below which repulsion is felt by particles. Suppose $\phi_{i}$ represents the net potential on the $i^{th}$ particle due to other particles, then $\phi_{i}=\sum\limits_{\substack{j}} V_{ij}(1-\delta_{ij})$. $\delta_{ij}$ is the standard kronecker delta function. The corresponding force felt is $ -\nabla_{\textbf{r}_{i}}\phi_{i}$. 

The confinement interaction is modeled by a piecewise harmonic repulsive potential. Suppose $b_{i} = \{x_{i},y_{i},z_{i}\}$ and $l_{b} = \{l_{x},l_{y},l_{z}\}$ constitute ordered pairs, where elements of $b_{i}$ represent cartesian components of $i^{th}$ particle's position and elements of $l_{b}$ represent edge-length of the box in $x$, $y$ and $z$ direction respectively. The confining potential corresponding to each component of $b_{i}$ and $l_{b}$ is denoted by $V_{b_{i}}$.

\begin{eqnarray}\nonumber
    V_{b_{i}} =
  \begin{cases}
                 \frac{1}{2}\kappa {b}_{i}^{2}& \text{if $ {b}_{i}<0,$} \\\\
                 0 & \text{if $ 0\leq b_{i} \leq l_{b}, $}\\\\
                \frac{1}{2}\kappa ( {b}_{i} - l_{b})^{2} & \text{if $ {b}_{i} > {l_{b}}$}.
  \end{cases}
\end{eqnarray}\\
The overall confining potential $V_{c} = \sum\limits_{\substack{ b_{i}}} V_{b_{i}} $. Thus, the net force felt by $i^{th}$ particle due to excluded volume and confinement effects is $\mathbf{F}_{i} = (-\nabla_{\textbf{r}_{i}}\phi_{i}) + (-\nabla_{\textbf{r}_{i}}V_{c})$. 

The coordinate-dependent damping is modeled by a hyperbolic tangent function in the simulations. The following functional form of damping experienced by $i^{th}$ particle, is used in simulations:
\begin{eqnarray*}
 \Gamma_{i}=A\hspace{1mm}\tanh{\beta(x_{i}-x_{0})}+ (A+1),
\end{eqnarray*}
where $x_{0}$ is shifted origin and $A$ is amplitude parameter, where $\beta$ is the measure of the steepness of the profile in neighborhood of $x_{0}$ and effectively is a measure of region $R_{2}$ in $x$ direction. The last term in the above expression of damping accounts for vertical shifting of standard hyperbolic tangent function to ensure that damping is always positive. We have chosen $x_{0}$ to be midpoint of length of three-dimensional box in $x$ direction.\\
\par{}
The initial configuration in case I corresponds to randomly placing all the particles in $R_{1}$. This initial placement sets up a particle concentration gradient that favors particle motion through $R_{2}$. Normal concepts of diffusion would dictate that the particles would diffuse to reach a uniform distribution. Coordinate dependence of damping sets up a diffusion potential according to the set diffusion gradient. This, when evolved as an It\^o-process, induces motion of particles in the direction of decreasing diffusivity (increasing damping) in accordance with the It\^o's framework. The combined result of the two processes is that the majority of particles diffuse to region $R_{3}$, and consequently, a steady state is reached with higher concentration in $R_{3}$.\\
\par{}
Now, consider case II, where all the particles are randomly placed in $R_{3}$ at the start of the simulation. This assignment creates a concentration gradient favoring the motion of particles through $R_{2}$ to $R_{1}$ for normal diffusion. However, the diffusion potential as per It\^o-process sets up a diffusion barrier to overcome to this concentration-driven current. Consequently, a steady state is reached at long times with most of the particles staying in $R_3$. This is a diode in reverse bias if one goes by the conventional idea of considering diffusion being driven by concentration gradient.\\ 
\par{}
 ${n}_{f}$ represents the ensemble-averaged number of particles reaching $R_{3}$ starting from $R_{1}$. It corresponds to case I. Similarly, ${n}_{b}$ denotes the ensemble-averaged number of particles occupying region $R_{3}$ given they start from $R_{3}$ and it corresponds to case II. It is evident from Fig.~\ref{fig_2} (a)(case I) that when particles are initially placed in $R_{1}$, the majority of them reach $R_{3}$ at long times. This is because the inhomogeneity-induced diffusion potential sets up a diffusion gradient which aids the concentration gradient in the motion in this (forward) direction. The situation is similar to forward biasing in a traditional diode by assistance in forward motion. In case II as evident from Fig.~\ref{fig_2} (b), most of the particles can't crossover to region $R_{2}$ because the inhomogeneity-induced diffusion potential generates a diffusion gradient that opposes the motion of particle from $R_{3}$ to $R_{2}$ (backward) direction. Simulation details used here can be found in Appendix A and Appendix B.  
 
\begin{figure}[h]
  \centering
  \subfloat[Case I ]{\includegraphics[width=0.90\textwidth]{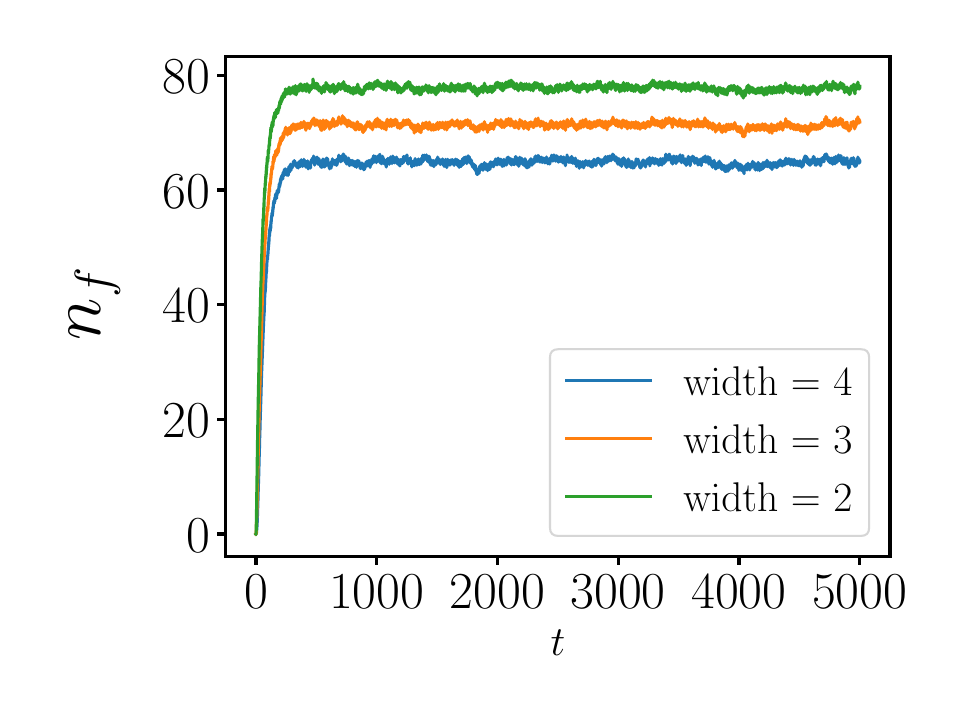}\label{fig:f1}}
  \hfill
  \subfloat[Case II ]{\includegraphics[width=0.90\textwidth]{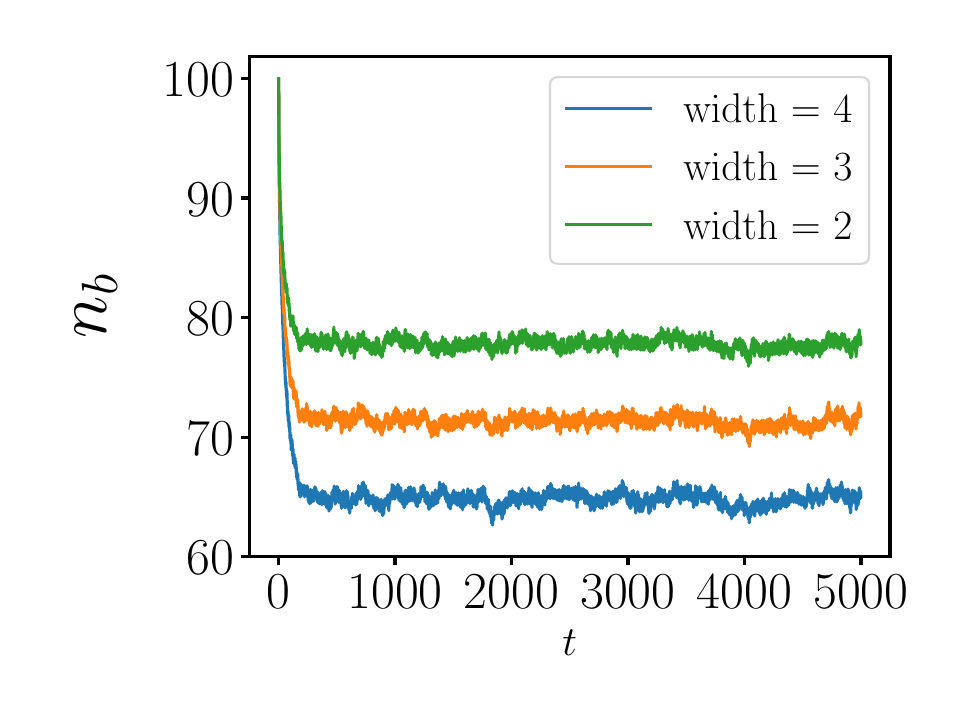}\label{fig:f2}}
  \caption { Occupation of particles in $R_{3}$, for various widths of $R_{2}$ as a function of time displayed for (a) case I and (b) case II. Here, widths of 4, 3, and 2 correspond to the following values of $\beta$: 2.5, 3.3, and 5 respectively.}
  \label{fig_2}
\end{figure}

\section{ Entropic pulling of a chain }

\par
In this section, we consider a polymer chain confined inside a three-dimensional funnel-like geometry in contact with a heat reservoir in thermal equilibrium at a temperature $T$. The polymer comprises $N$ interacting Brownian monomers. The shape of the funnel is shown in Fig.~\ref{fig_3}. It is useful to think of the funnel as divided into three sections: $R_{1}$, $R_{2}$, and $R_{3}$, all seamlessly linked together. $R_{1}$ is a hollow right circular cylindrical region of radius $\mathcal{R}$. $R_{2}$ is a hollow truncated cone-shaped region (or frustum of a cone) with big radius $\mathcal{R}$ and small radius $r$. $R_{3}$ is a hollow right circular cylindrical region of radius $r$. All regions have equal length by construction. We establish our reference point, the origin, at the center of the rear circular base (radius $\mathcal{R}$) of the funnel, which marks the beginning of region $R_{1}$. The positive direction of $z$ axis is a ray coincident with the axis of symmetry of funnel emerging from origin towards other regions, with $x$ and $y$ axis perpendicular to $z$ axis. It is evident from the nature of geometry of the funnel that curved boundaries of regions $R_{1}$, $R_{3}$ lie at a constant radial distance from $z$ axis, whereas for $R_{2}$, this distance keeps on decreasing with increasing $z$ coordinate. Thus, this geometry can be used to model the effect of the wall on diffusivity/damping on the particle.\\
\par{}
We introduce a confining potential on boundaries to ensure that polymer stays inside the funnel. In the model, we introduce two intra-polymer interactions: (i) nearest-neighbor monomer interaction and (ii) interaction between all other pairs. The nearest neighbor monomers feel attractive interaction above an equilibrium separation and an excluded volume interaction below it. All other pairs of monomers interact via excluded volume interaction in proximity to one another. We evolve the system as an It\^o process. The following equation of motion describes the dynamics of $i$-th monomer under It\^o description.

\begin{align*}
\begin{split}
\frac{  d{ \mathbf { r}_{i}  } }{dt} &= -\frac{1}{\Gamma_{i}}\nabla_{\mathbf{r}_{i}} \Big[ \sum\limits_{\substack{ j = i\pm 1}}V_{NN}(|\mathbf{r}_{ij}|)+\sum\limits_{\substack{j\\ j\neq i,i\pm 1}}  V_{O}(|\mathbf{r}_{ij}|)\Big ]\\
&-\frac{1}{\Gamma_{i}}\nabla_{\mathbf{r}_{i}} V_{B}(\mathbf{r}_{i} )+ \sqrt\frac{2k_{B}T}{ \Gamma_{i}} {\boldsymbol{\eta}}_{i}(t).
\end{split}
\end{align*}

where $i,j $ $\in$ $\{1,2,\ldots\,N\}$ are monomer indices and  $\mathbf {r}_{i}$ denotes the position vector of $i^{th}$ monomer. ${r}_{ij}=|\mathbf{r}_{ij}|=|\mathbf{r}_{i}-\mathbf{r}_{j}|$ is the magnitude of separation between  $i^{th}$ and $j^{th}$ monomer. $\Gamma_{i}$ is the damping coefficient associated with the $i^{th}$ monomer. $V_{NN}$ captures the interaction between monomers directly linked to one other, while $V_{O}$ accounts for the interaction between monomer that are not nearest neighbour pairs. $V_{B}$ denotes the potential that restrains the $i$-th monomer in the event it crosses funnel boundary. ${\boldsymbol{\eta}}_{i}(t)$ signifies Gaussian vector white noise experienced by $i$-th monomer possessing same statistical properties as those elucidated  in the {\it diffusion diode section}.

\begin{figure}[h!tbp]
\includegraphics[scale=0.16]{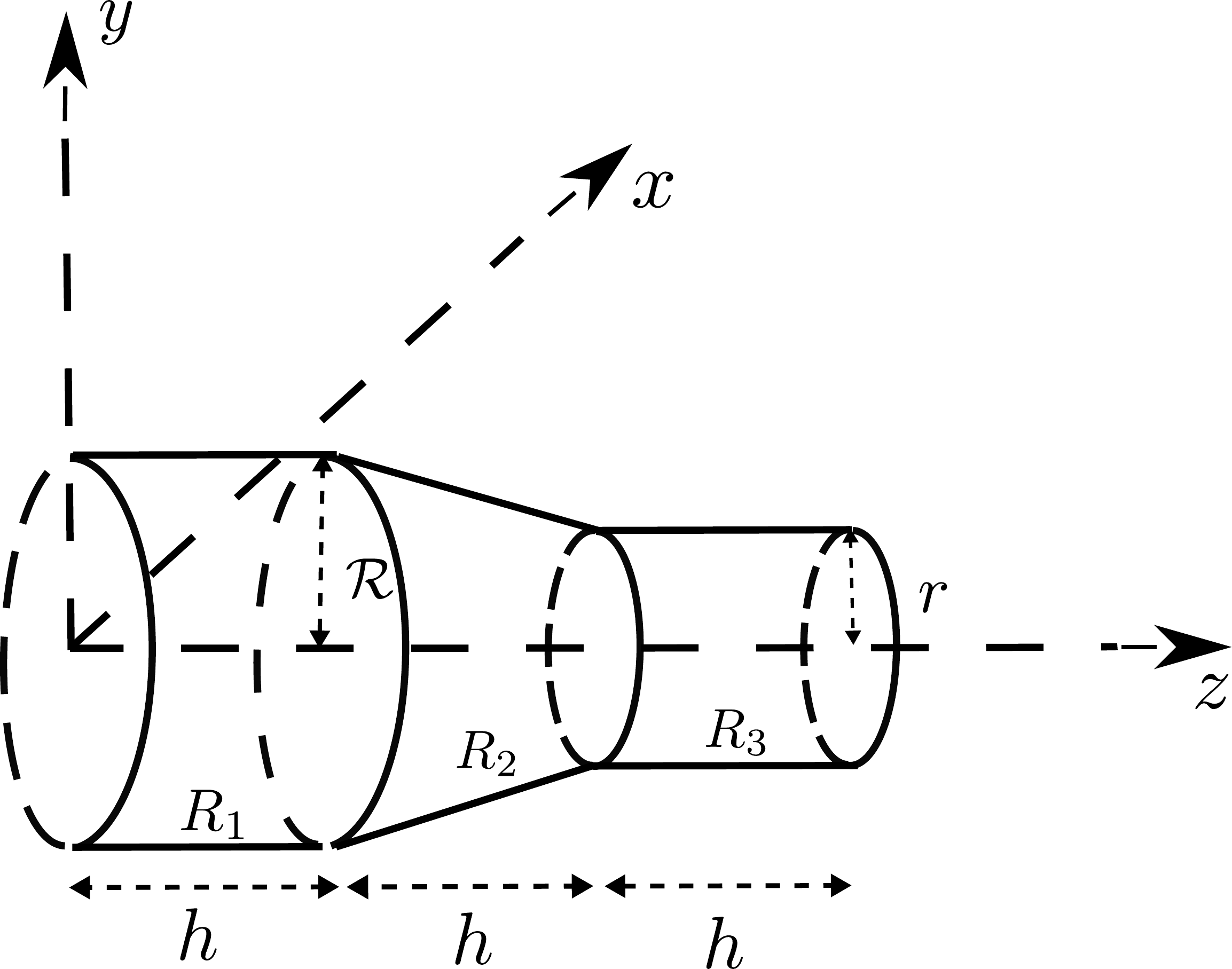}
\caption{ \footnotesize  Funnel (not drawn to scale) }
  \label{fig_3}
\end{figure}

The cross-sectional radius $s$ in each region of the funnel can be expressed as:
  \begin{equation*}
  s(z) =
  \begin{cases}
                 \mathcal{R} & \text{if $ 0 \le z \le h,$} \\\\
                {r +(2h-z)(\frac{\mathcal{R}-r}{h})}  & \text{if $ h \le z \le 2h, $} \\\\
                r  & \text{if $  2h \le z \le 3h, $}
  \end{cases}
\end{equation*}\\
where $h$ is each region's length in $z$ direction. The nearest neighbor monomer interaction $V_{NN}$ is modeled by a harmonic potential. 
\begin{equation*}
V_{NN}({r}_{ij}) = \frac{\alpha}{2} ({r}_{ij}- r_{min})^2 ,   
\end{equation*}
where $\alpha$ is the spring constant, $r_{min}$ is the equilibrium separation and $r_{ij}$ is the separation between nearest neighbour's $i$ and $j$, where ($j = i \pm 1$). The excluded volume interaction between the rest of  monomer pairs is modeled using {\it Weeks–Chandler–Andersen (WCA)} potential and represented by $V_{O}$.

\begin{equation*}
V_{O}({r}_{ij}) =
  \begin{cases}
                 4\epsilon[\frac{\sigma^{12}}{{r}_{ij}^{12}}-\frac{\sigma^{6}}{{r}_{ij}^{6}}] + \epsilon & \text{if $  {r}_{ij}<2^{\frac{1}{6}}\sigma,$} \\\\
                0  & \text{if ${r}_{ij} \geq 2^{\frac{1}{6}}\sigma, $}
  \end{cases}
\end{equation*}
where, $\epsilon$ measures the depth of the standard {\it L-J potential}  well, $\sigma$ is a length parameter and $r_{ij}$ is the separation between $i^{th}$ and $j^{th}$ monomer with ${ j\neq i,i\pm 1}$. \\
\par{}
 
The position of $i^{th}$ monomer in three-dimensional space can always be expressed in a cartesian coordinate system by three independent cartesian coordinates, \{$\mathbf{r}_{i}$\} $\Longleftrightarrow$ \{$x_{i}$, $y_{i}$, $z_{i}$\}. The presence of cylindrical symmetry in the funnel allows for a convenient representation of confining potential at the periphery and above it, in cylindrical coordinates, \{$\mathbf{r}_{i}$\} $\Longleftrightarrow$ \{$\rho_{i}$, $\phi_{i}$, $z_{i}$\}. $\rho_{i}$ symbolises the radial distance of the $i^{th}$ monomer from $z$ axis and $\phi_{i}$ expresses the azimuth angle made by $i^{th}$ monomer on $x$-$y$ plane, where $\rho_{i} = \sqrt{x_{i}^{2}+y_{i}^{2}}$ and $\phi_{i} =\tan^{-1}(\frac{y_{i}}{x_{i}})$. The funnel boundary consists of two types of surfaces: a curved cylindrical surface and a plane circular surface towards the rear ends of the funnel along the $z$ axis. The confinement interaction is defined separately for both surfaces but in either case, is modeled by harmonic repulsion. Suppose $\Phi_{c}$ and $\Phi_{p}$ signify the confining potentials that the  $i^{th}$ monomer experiences on overshooting the curved cylindrical boundary and plane circular periphery of the funnel, respectively.\\

\begin{align*}
\Phi_{c}(\rho_{i}) &=
\begin{cases}
0 & \text{if } \rho_{i}<s_{i}, \\ \\
\frac{\kappa}{2} (\rho_{i}- s_{i})^2 & \text{if } \rho_{i} \geq s_{i}.
\end{cases}
\\ \\
\Phi_{p}(z_{i}) &=
\begin{cases}
\frac{\kappa}{2} (z_{i}-0)^2 & \text{if } z_{i}\leq 0, \\ \\
0 & \text{if } 0<z_{i} < 3h, \\ \\
\frac{\kappa}{2} (z_{i}-3h)^2 & \text{if } z_{i} \geq 3h.
\end{cases} \\
\end{align*}

where, $\kappa$ is measure of strength of harmonic repulsion, $s_{i}(z_{i})$ is the cross-sectional radius seen by $i^{th}$ particle corresponding to it's $z$ coordinate, $z_{i}$. Thus, the overall confining potential $V_{B}$ becomes, $V_{B}$ = $\Phi_{c} +\Phi_{p}$. 
\par{}
Having described the system, in the following section, we show the effect of the presence and absence of coordinate-dependent damping on the dynamics of the polymer chain. We consider two cases. Case I corresponds to the introduction of damping dependent on $z$ coordinate,  $\Gamma_{i} = A/s_{i}(z_{i})$, where $A$ is a constant. Such a choice provides uniform damping in regions $R_{1}$ and $R_{3}$ and a monotonically increasing damping as the polymer moves in region $R_{2}$ along increasing $z$ direction.  In Case II, we consider a constant damping profile experienced by the monomer throughout the funnel, $\Gamma_{i} = A/R$, which equals a constant. We evolve both cases as an It\^o  process using simulation and compare the dynamics at the end of the simulation.

\begin{figure}[t]
  \centering
  \subfloat[]{\includegraphics[width=0.90\textwidth]{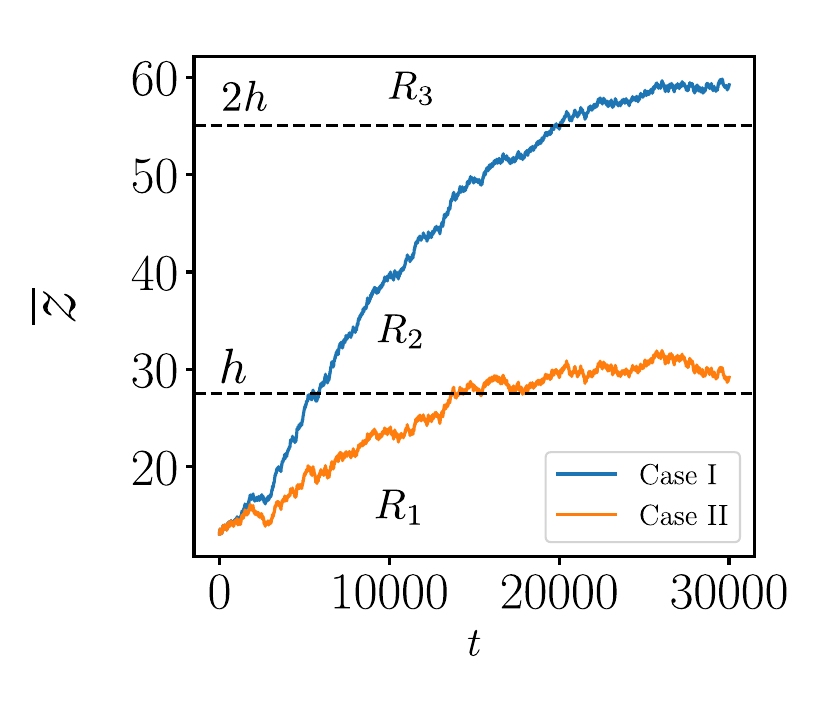}\label{fig:f1}}
  \hfill
  \subfloat[ ]{\includegraphics[width=0.90\textwidth]{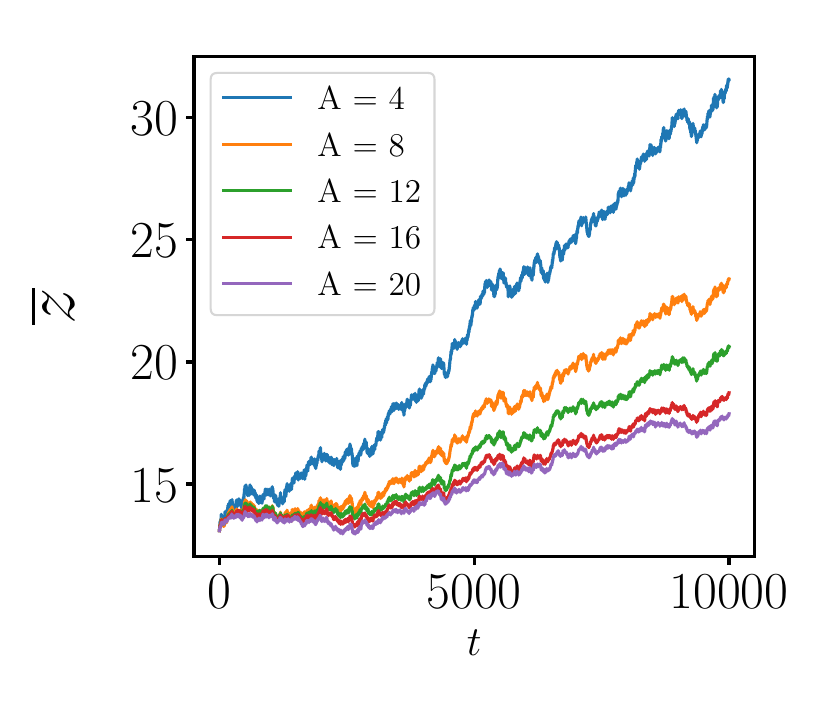}\label{fig:f2}}
  \caption { Time evolution of the center of mass of 25 member polymer chain, $\Bar{z}$ (a) compared in the presence (case I) and absence (case II) of coordinate dependent damping; (b) compared in the presence of coordinate dependent damping for various damping strengths as shown.}
  \label{fig_4}
\end{figure}

\par{}
The initial state of the polymer in cases I and II is an identical unfolded linear chain conformation coincident on the $z$ axis lying in $R_{1}$ in all the ensembles. We generate independent ensembles using a different random seed and thus a different sequence of random numbers for each ensemble in case I. For case II, we generate an equal number of independent ensembles by using the same random seed and the same corresponding sequence of random numbers as for case I.
\par{}
After having fixed the polymer to the same initial configuration, and the same sequence of random numbers corresponding to a given ensemble in both cases, we can distinctly see the effect of the presence of coordinate-dependent damping in case I and compare it with constant damping (absence of coordinate dependence) as in case II. We examine this by comparing the dynamics of ensemble-averaged $z$ coordinate center of mass of the polymer denoted by $ \Bar{z}$.
\par{}
It is evident from Fig.~\ref{fig_4}(a), that due to the presence of coordinate-dependent damping (case I), the polymer moves from a wider region of $R_{2}$ to its narrower region. The presence of coordinate-dependent damping sets up an entropy gradient that pulls the polymer from the wider end to the narrow end of $R_{2}$, ending up in $R_{3}$. The polymer stays in $R_{3}$ due to the inability to move against the entropy gradient present in $R_{2}$. However, in the uniform damping case (case II), there exists no such entropy gradient, so the polymer keeps diffusing around in the wider region. The movie of the time evolution of a particular realization comparing the effect of presence and absence of coordinate-dependent damping can be found \href{https://youtu.be/TQU67Cjg5qo} {here}.\par{}
Now, we remove regions $R_{1}$ and $R_{3}$ and place the Brownian polymer chain in a geometry that consists of only region $R_{2}$ in thermal equilibrium with a heat bath at temperature $T$. The confinement potential is kept intact. The damping is coordinate dependent, $\Gamma_{i} = \frac{A}{s_{i}(z_{i})}$, where $s_{i}(z_{i}) = r +(h-z_{i})(\frac{\mathcal{R}-r}{h})$. The polymer is initialized in an unfolded linear configuration along the $z$ axis. We study the effect of variation of strength of diffusive gradient on transport of the polymer. We systematically increase $A$, run the simulations and then compare the dynamics of the ensemble-averaged center of mass of the polymer in $z$ direction. As, the velocity (current) $\overline{v_{z}}$ attained (in $z$ direction)is proportional to the gradient of diffusivity, a larger $A$, generates a steeper damping gradient which corresponds to smaller diffusivity gradient, hence weaker transport. This is clearly visible in Fig.~\ref{fig_4}(b). Fig.~\ref{fig_5} shows the variation of velocity of the center of mass of polymer in $z$ direction, $\overline{v_{z}}$  as function of diffusivity strength. To verify, the linear relationship between $\overline{v_{z}}$  and $A^{-1}$ (measure of diffusivity strength), we performed a linear fit and found slope of $0.0068$, intercept of $0.00016$ and $R^{2}$ of 0.999. Thus, velocity is proportional to measure of diffusivity strength (inverse damping strength), $\frac{1}{A}$. Simulation details used here can be found in Appendix A and Appendix C.

\begin{figure}[t!tbp]
  \centering
  {\includegraphics[width=0.90\textwidth]{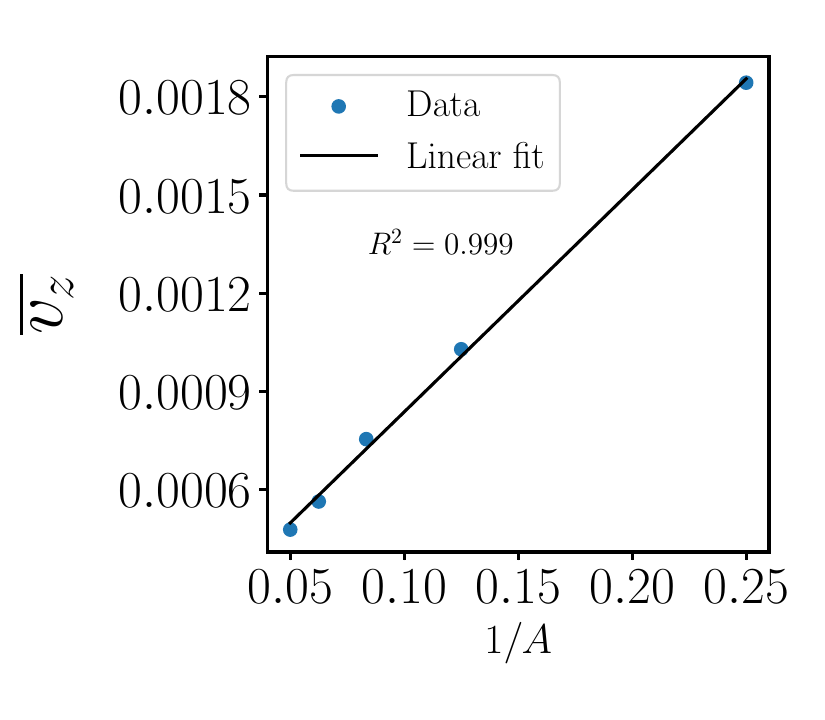}\label{fig:f1}}
  \caption {Variation of Center of mass velocity of polymer vs inverse damping strength $A^{-1}$ shown by data points and a linear fit is performed. The linear fit correspond to slope of $0.0068$, intercept of $0.00016$ and $R^{2} =0.999$. }
  \label{fig_5}
\end{figure}

\section{Discussion}

People consider it to be counter-intuitive in the context of diffusion to think of diffusion creating concentration gradient. However, an It\^o process demonstrates that it is counter-intuitive in the uniform diffusion. It turns out to be a general phenomenon in the presence of diffusion gradients when the process is treated as an It\^o-process. An It\^o-process does not introduce correlations in the thermal noise and the thermal noise remains a white noise. Note that, any coloured noise in the over-damped dynamics will destroy the local thermal equilibrium of a Brownian particle by violating the second fluctuation-dissipation relation. The distribution that results from an It\^o process has two factors, the canonical Boltzmann factor and a micro-canonical density of states arising from the inhomogeneity of space due to coordinate dependence of diffusion. Therefore, structurally, the distribution is a legitimate distribution which has been arrived at from a stochastic process which has not considered any temporal correlations in the noise.

The It\^o-process of coordinate-dependent diffusion results in striking consequences by explicitly keeping in place the effects of existing diffusion gradients in the distribution. It primarily results into the acceptance of the interplay between two possible gradients namely the concentration gradient and the diffusion gradient. The possibility of this competition between two gradients when diffusion is coordinate dependent is never explicitly taken into account in the existing literature when one does not consider the phenomenon as an It\^o process. A whole lot of situations can arise under the existence of this interplay of gradients of which the effective attraction of a particle from a wider region to a narrower region is demonstrated in this paper which provides a physical basis for sought after entropic pulling mechanism in biological systems. Obviously, observation in those biological systems had indicated the need for the hypothesis of the entropic pulling which could not be based upon the phenomenon of uniform diffusion. Coordinate-dependent diffusion, with all its controversy in relation to identifying the process to be It\^o or Stratonovich or something else, provides a definite hint of existence of such an entropic pulling mechanism when the process is an It\^o process. 
 
\begin{acknowledgments}
 Mayank acknowledges National Supercomputing Mission for the use of PARAM Brahma at IISER Pune. Mayank also acknowledges financial support from the CSIR-SRF-Direct fellowship. Mayank would like to acknowledge Ritam Pal and Hitendra for their help and inputs in data analysis and video animation. Mayank also acknowledges fruitful discussions on biological systems with Prajna Nayak.
\end{acknowledgments}

\section*{Appendix A : BASICS OF AN IT\^O PROCESS}
Consider the following overdamped Langevin equation for the $i^{th}$ particle in presence of a force $\mathbf{F}_{i}$, Gaussian noise $\boldsymbol{\eta}_{i}(t)$ and coordinate-dependent damping $\Gamma_{i}$ in equilibrium with heat bath at temperature $T$:
\begin{eqnarray}{\nonumber}
    \frac{  d{ \mathbf { r}_{i} } }{dt} &=& \frac{ \mathbf{F}_{i}}{\Gamma_{i} } +\sqrt\frac{2k_{B}T}{ \Gamma_{i} } {\boldsymbol{\eta}}_{i}(t).
\end{eqnarray}

  The equation in this paper is numerically evolved using the Euler-Maruyama scheme and by considering it as an It\^o process. The It\^o process corresponds to utilizing a correlation-free (non-anticipating) noise. This means while numerically evolving the dynamics, at every time step ($\Delta t$) of evolution, the particle's position at the start of the interval $\Delta t$ determines the damping/diffusivity. Thus when considered  as an It\^o process, the following Euler-Maruyama updation prescription is used:

\begin{eqnarray}{\nonumber}
\mathbf{r}_{i}(t + \Delta t) &= \mathbf{r}_{i}(t)& +\frac{ \mathbf{F}_{i} \{\mathbf{r}_{i}(t)\} }{\Gamma_{i}  \{\mathbf{r}_{i}(t)\}} \Delta t +\sqrt\frac{2k_{B}T \Delta t}{ \Gamma_{i} \{\mathbf{r}_{i}(t)\}} \boldsymbol{\mathcal{N}}(0,1),
\end{eqnarray}

Here $\Delta t$ is the discretized time step by which particle's position is updated after each iteration. $\mathbf{F}_{i} \{\mathbf{r}_{i}(t)\}$ and $\Gamma_{i}  \{\mathbf{r}_{i}(t)\}$  respectively denote the net force and damping coefficient of $i^{th}$ particle at time $t$. $\boldsymbol{\mathcal{N}}(0,1)$ is a three-dimensional Gaussian distributed random variable vector with zero mean, unit standard deviation, and zero cross-correlation. \\

We implement the following methodology for our simulations:
\begin{enumerate}
\item[(1)] Initialise the system.
\item[(2)]Evaluate the net force and damping associated with each particle. 
\item[(3)]Generate  Gaussian distributed random variables with the above mentioned desirable properties and then update the position of each particle using the It\^o process discretization prescription.
\item[(4)]  Go back to step (2) and repeat.
\end{enumerate}

\section*{ Appendix B : { Diffusion diode}}

Initialization in the {\it Diffusion diode} section corresponds to assigning random positions to particles inside their respective regions ($R_{1}$  and $R_{3}$ for cases I and II, respectively) at the start of the simulation. In the simulations, new random seeds and different uncorrelated sequences of random numbers are used to generate Gaussian-distributed random variables. We run the simulation till a steady state is reached. Thermal energy $k_{B}T$ and distance parameter $r_{l}$ are our reference scales for energy and length in simulation.
\par{}
 We fix $k_{B}T = 1$, $r_{l} = 0.25$ and number of particles $N=100$, with repulsion strength parameter set to $\kappa =1000$. The box dimensions chosen are $l_{x}=14$, $l_{y}=l_{z} =5$. To model the damping profile, we chose $A = 4.5$ and $x_{0} = \frac{1}{2}l_{x} =7$. We use discretized time step $\Delta t = 5 \times 10^{-4}$ to evolve equation $(3)$. We consider three different damping profile parameters, which corresponds to changing $\beta$ keeping other damping parameters $A$ and $x_{0}$ fixed. We simulate this for both cases (I and II). The choice of $\beta$ decides the width of $R_{2}$ in $x$ direction. The following values of $\beta$'s are chosen: 2.5, 3.3, and 5.0, corresponding to $R_{2}$'s  widths (in $x$ direction) 4, 3, and 2 respectively. Each simulation is run for $10^{7}$ iterations with successive data recorded every $100$ iterations. We generate $100$ ensembles for each case for a fixed $\beta$ and perform an ensemble average to obtain quantities of interest. ${n}_{f}$ and ${n}_{b}$ are the quantities of interest here, obtained by averaging over 100 ensembles.\\

\section*{APPENDIX C : Entropic pulling of a chain}
 The polymer is initialized in an identical unfolded linear chain conformation (coincident on $z$ axis lying in $R_{1}$ at the start of the simulation) in all the ensembles for cases I and II. We generate $50$ ensembles for cases I and II. $k_{B}T$ and $r_{min}$ serve as fundamental energy and length benchmarks in our simulations. Both are normalized to unity, $k_{B}T = 1$ and $r_{min} = 1$. The polymer consists of $N=25$ monomers. The geometric parameters defining the funnel geometry are $\mathcal{R}=3$, $r=2$, $h = 27.5$. Interaction parameters are set at $\alpha =100$, $\epsilon = 1.5$, $\sigma = 1.5$, $\kappa =1000$ and $A=4$. Discretised time step $\Delta t = 10^{-4}$ is employed and the simulations is run for $3\times 10^{8}$ iterations with monomer positions recorded every $10^{3}$ iterations. The video at the end of discussion on Fig.~\ref{fig_4}(a) for each case is prepared using matplotlib library, with frame rate of $1000$ and an interval of $1$ millisecond between two consecutive frames. It is then integrated into single place using a video editor.\\
Now when we remove regions $R_{1}$ and $R_{3}$ and place the polymer chain in a geometry that only consists of $R_{2}$, we use following parameters: ${R}=3$, $r=2$, $h = 200$, interaction parameters are set at $\alpha =100$, $\epsilon = 1.5$, $\sigma = 1.5$, $\kappa =1000$, and time step of $\Delta t = 10^{-4}$ is employed. The damping is coordinate dependent, $\Gamma_{i} = \frac{A}{s_{i}(z_{i})}$, where $s_{i}(z_{i}) = r +(h-z_{i})(\frac{\mathcal{R}-r}{h})$. The simulations is run for $ 10^{8}$ iterations with monomer positions recorded every $10^{3}$ iterations. We generate 200 ensembles for each value of $A$ and evaluate $\Bar{z}$ and $\overline{v_{z}}$.\\

\nocite{*}
\bibliography{fileone}

\newpage
\onecolumngrid

 \end{document}